\documentclass[runningheads]{llncs}
\usepackage[T1]{fontenc}
\usepackage{graphicx}
\usepackage[hidelinks]{hyperref}
\usepackage{color}

\usepackage{listings}
\usepackage{tikz}
\usepackage{adjustbox}
\usepackage{enumitem}

\usetikzlibrary{automata,positioning}

\lstdefinelanguage[RISCV]{Assembler}
{
  alsoletter={.}, 
  alsodigit={0x}, 
  morekeywords=[1]{ 
    lb, lh, lw, lbu, lhu, mv,
    sb, sh, sw, mul, addi, li,
    sll, slli, srl, srli, sra, srai,
    add, addi, sub, lui, auipc,
    xor, xori, or, ori, and, andi,
    slt, slti, sltu, sltiu,
    beq, bne, blt, bge, bltu, bgeu,
    j, jr, jal, jalr, ret,
    scall, break, nop
  },
  morekeywords=[2]{ 
    .align, .ascii, .asciiz, .byte, .data, .double, .extern,
    .float, .globl, .half, .kdata, .ktext, .set, .space, .text, .word
  },
  morekeywords=[3]{ 
    zero, ra, sp, gp, tp, s0, fp,
    t0, t1, t2, t3, t4, t5, t6,
    s1, s2, s3, s4, s5, s6, s7, s8, s9, s10, s11,
    a0, a1, a2, a3, a4, a5, a6, a7,
    ft0, ft1, ft2, ft3, ft4, ft5, ft6, ft7,
    fs0, fs1, fs2, fs3, fs4, fs5, fs6, fs7, fs8, fs9, fs10, fs11,
    fa0, fa1, fa2, fa3, fa4, fa5, fa6, fa7
  },
  morecomment=[l]{;},   
  morecomment=[l]{\#},  
  morestring=[b]",      
  morestring=[b]'       
}

\lstdefinelanguage{SML}
{
  morekeywords= {
    Definition, Theorem, Proof, QED, End, EQUAL, GREATER, LESS, NONE, SOME, abstraction, abstype, and, andalso, array, as, before, bool, case, char, datatype, do, else, end, eqtype, exception, exn, false, fn, fun, functor, handle, if, in, include, infix, infixr, int, let, list, local, nil, nonfix, not, o, of, op, open, option, orelse, overload, print, raise, real, rec, ref, sharing, sig, signature, string, struct, structure, substring, then, true, type, unit, val, vector, where, while, with, withtype, word
  },
  morestring=[b]",
  morecomment=[s]{(*}{*)},
  keywordstyle=\bfseries
}

\begin{document}
\title{Trustworthy Verification of RISC-V Binaries Using Symbolic Execution in HolBA}

\titlerunning{Trustworthy Verification of RISC-V Binaries}

%
\author{Karl Palmskog\inst{1} \and 
Andreas Lindner\inst{2} \and
Scott Constable\inst{3} \and
Roberto Guanciale\inst{1} \and
Hamed Nemati\inst{1}}
\authorrunning{K. Palmskog et al.}
%
\institute{KTH Royal Institute of Technology, Stockholm, Sweden\\
\email{\{palmskog,robertog,hnemati\}@kth.se}\\ \and
Uppsala University, Uppsala, Sweden\\
\email{andreas.lindner@angstrom.uu.se}\\ \and
Intel Labs, USA\\
\email{scott.d.constable@intel.com}}

\maketitle              
\begin{abstract}
Many types of formal verification establish properties about abstract high-level program representations, leaving a large gap to programs at runtime. Although gaps can sometimes be narrowed by techniques such as refinement, a verified program's trusted computing base may still include compilers and inlined assembly.
In contrast, verification of binaries following an Instruction Set Architecture (ISA) such as RISC-V can ensure that machine code behaves as expected on real hardware.
While binary analysis is useful and sometimes even necessary for ensuring trustworthiness of software systems, existing tools do not have a formal foundation or lack automation for verification. 
We present a workflow and toolchain based on the HOL4 theorem prover and the HolBA binary analysis library for trustworthy formal verification of RISC-V binaries. The toolchain automates proofs of binary contracts by forward symbolic execution of programs in HolBA's intermediate language, BIR.
We validated our toolchain by verifying correctness of RISC-V binaries with (1)~an implementation of the ChaCha20 stream cipher and (2)~hand-written assembly for context switching in an operating system kernel.


\end{abstract}

\section{Introduction}
%
Many types of formal verification are based on using an abstract representation of a program, e.g., a pure function or high-level program syntax inside an interactive theorem prover (ITP). While this approach may leave a large gap 
between the verified abstraction and the actual code being executed, engineers can use techniques such as step-wise refinement or extraction to extend verification down to executable source code~\cite{Cohen2013}. Still, the Trusted Computing Base (TCB) for such verified programs could include unverified compilers and inlined hand-written assembly, which may undermine formal guarantees.

In contrast to high-level source code, machine code (assembly) in binary format is what is typically executed directly on the underlying hardware, and Instruction Set Architectures (ISAs) such as RISC-V~\cite{Waterman2016} have a well-defined machine state and a limited number of simple operations. Hence, by targeting this lower abstraction level, \emph{binary analysis tools}~\cite{bap,bitblaze} can ensure strong properties, e.g., by ruling out issues with compilers and inlined assembly. 
%
However, while binary analysis is useful---and even necessary for ISA-level guarantees in software systems tied to a proprietary unverified compiler---existing tools either do not have a formal foundation or lack automation for verifying non-trivial high assurance machine code, such as for cryptographic ciphers.

We present a toolchain based on the HOL4 theorem prover~\cite{hol4} and the HolBA library~\cite{lindner201972} for trustworthy formal verification of RISC-V (RV64G instruction set) binaries. The toolchain is based on \emph{lifting} RISC-V disassembly to the intermediate language BIR embedded in HOL4 and then performing proof-producing \emph{forward symbolic execution} of BIR programs. We express properties of binaries to be verified using binary-level contracts~\cite{Lundberg2020} via the independently developed L3 model of RISC-V~\cite{Fox2015}; when such contracts are translated to BIR level and proved automatically using symbolic analysis, we \emph{backlift} the results to obtain RISC-V contract theorems about lifted binaries.

Ultimately, our goal is to automate key steps of \emph{translation validation}, where binary code is demonstrated to refine the behavior of an abstract algorithmic specification~\cite{Sewell2013}. Intuitively, this means that our toolchain is not meant for proving that some assembly routine sorts a list, but instead that it refines (is simulated by) a specific merge sort implementation. Hence, to validate our toolchain, we apply it to case studies from two relevant high-assurance domains: cryptography and operating system kernels.
As with most ITP-based tools, some parts in our toolchain are fully automated (albeit incomplete), while others require manual interaction with HOL4, e.g., to express specifications. 


In summary, we make the following contributions:
\begin{enumerate}
\item We describe a workflow for verifying RISC-V binaries, which we implemented by way of a toolchain on top of HOL4 and HolBA using proof-producing symbolic execution.
\item We demonstrate the toolchain in two case studies: one binary implementing the ChaCha20 stream cipher~\cite{Bernstein2008}, and one binary with hand-written RISC-V assembly from a separation kernel~\cite{s3k}.
\item We perform an evaluation of our symbolic execution approach to investigate whether it scales to large RISC-V binaries.
\end{enumerate}
The toolchain components enabling our workflow are all included in the HolBA library and available together with examples under an open source license~\cite{holbasymbexec}.




\section{Background and Related Work}
%
In this section, we give some background on binary analysis, and contrast our binary verification with related approaches and previous work.

\paragraph{Binary analysis and intermediate languages.}

\begin{figure}[t]
\centering
\includegraphics[width=0.65\textwidth]{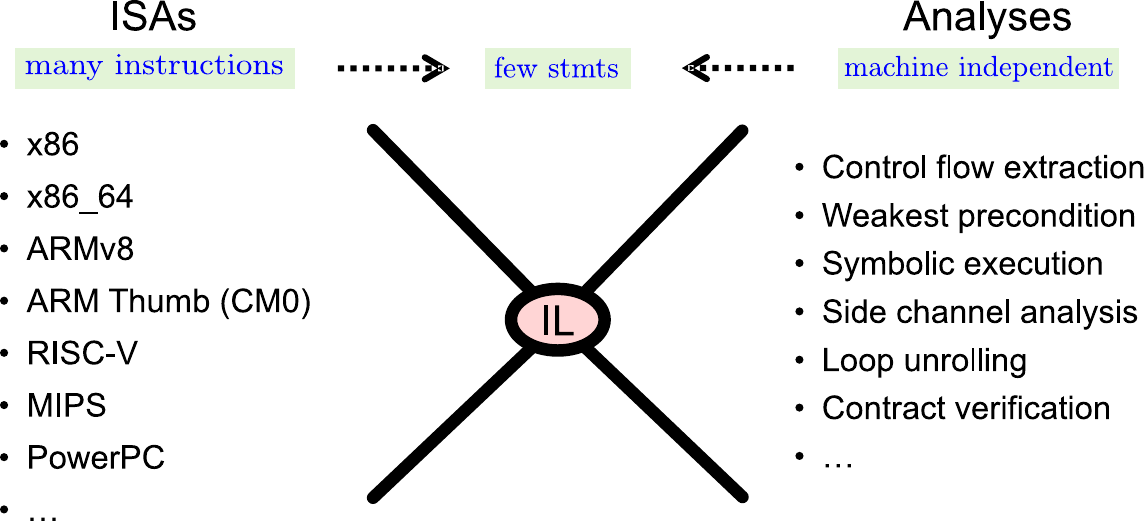}
\caption{Binary analysis tool architecture centering on an intermediate language (IL).} \label{fig:binalys}
\end{figure}

Analysis of machine code independent of any high-level language goes back to the 1940s~\cite{goldstine1947planning}, but received little attention in the context of formal verification until the late 1980s~\cite{Clutterbuck1988}.
%
Modern binary analysis tools include BitBlaze~\cite{bitblaze} and the Binary Analysis Platform (BAP)~\cite{bap}. These and similar tools, including HolBA, use the architecture shown in Figure~\ref{fig:binalys}, where binaries based on many different ISAs are translated to a single well-defined intermediate language (IL). This IL, called BIL for BAP and BIR for HolBA, is then used for many different analyses, e.g., symbolic execution and control flow, with results expressed in terms of the corresponding binary.


%
%
\paragraph{Binary verification and verified compilation.}
Properties of program source code can be formally transferred to a corresponding binary by a \emph{verified compiler} such as CompCert~\cite{Leroy2009} and CakeML~\cite{Tan2019}. Verified compilation (VC) is an orthogonal approach to direct binary verification (BV), with different advantages and limitations. Most directly, VC requires access to program source code, while BV does not. ITP-based environments for program verification based on VC, such as VST~\cite{Cao2018} and Characteristic Formulae~\cite{Gueneau2017}, do not allow expression of ISA-level properties, while BV permits (only) such properties. Inlined assembly in source code must be explicitly supported for VC, while BV is unaffected. 
Finally, support for custom ISA extensions, e.g., for hardware acceleration of cryptographic code, may require extensive additions to the verified compiler and its correctness theorem. In contrast, BV support for ISA extensions typically only requires minor changes to the IL and the process of converting binaries to the IL.

Binary verification can also be achieved indirectly by decompiling the binary into a model that resembles the source language (e.g., C) and then verifying that model. This approach was used to extend the seL4 refinement proofs from C source code to the compiled binary~\cite{Sewell2013}. The approach's coverage is limited by the decompiler, which may be unable to produce models for unconventional control flow behaviors or for ISA extensions that are not captured by the modeling language. For example, the decompiler used by Sewell~et~al.~\cite{Sewell2013} was unable to generate models for several procedures that originated from assembly source.



\paragraph{The HolBA library.}
HolBA is a library for binary analysis based on the HOL4 theorem prover~\cite{Slind2008}. Initial work on HolBA defined the BIR intermediate language for low-level code and functionality for proof-producing \emph{lifting} of binaries~\cite{lindner201972}. Subsequent work integrated a Hoare-style logic for expressing and reasoning about properties of BIR programs and for ISAs supported by HolBA~\cite{Lundberg2020}.
However, this previous work did not include domain-specific \emph{automation} for verification of binary programs. That is, verification had to be done by manually applying Hoare-style rules inside HOL4. Moreover, before the work described in this paper, end-to-end verification of RISC-V binaries was not possible.


\section{Binary Verification Workflow}
\begin{figure}[t]
\centering
\includegraphics[width=0.7\textwidth]{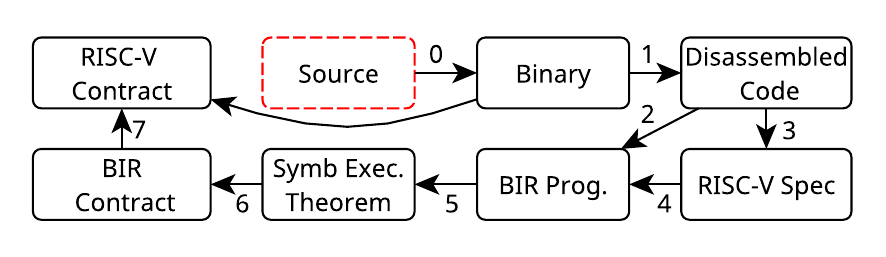}
\caption{Binary analysis workflow using HOL4 and HolBA.} 
\label{fig:workflow}
\end{figure}

%
The intended workflow of our toolchain consists of the steps below and is illustrated in Figure~\ref{fig:workflow}. While other workflows are possible using HolBA, e.g., for other ISAs such as ARMv8, we focus on a highly automated workflow for verification of functional correctness properties for RISC-V. As with most verification approaches using ITPs, our workflow is \emph{incomplete} in the sense that verification steps can fail without a specification being false.


\begin{enumerate}[font=\bfseries,start=0]
\item \textbf{Compilation:} Compile a source program to a RISC-V binary. This step is \emph{completely optional}, but adjustments to compilation can simplify verification. 
\item \textbf{Disassembly:} Disassemble the RV64G binary.
\item \textbf{Lifting:} Translate the disassembly to BIR and generate a HOL4 theorem connecting the BIR program to corresponding binary code.
\item \textbf{RISC-V specification:} Specify program boundaries inside disassembly and define pre- and post-conditions using the L3 model of RISC-V.
\item \textbf{BIR specification:} Translate RISC-V pre-conditions and post-conditions to BIR expressions and prove equivalence in HOL4.
\item \textbf{Symbolic execution:} Symbolically execute the BIR program according to the program boundaries and obtain a HOL4 symbolic analysis theorem.
\item \textbf{Symbolic transfer:} Prove a HOL4 contract theorem for the BIR program using the pre- and post-conditions via the symbolic analysis theorem. 
\item \textbf{Backlifting:} Prove a RISC-V contract theorem by way of the BIR contract theorem and the lifting theorem.
\end{enumerate}











%
To explain each of the steps, we use the running example defined in Figure~\ref{fig:ex-prog}, which shows different representations of a simple program \texttt{incr} that increments its argument (a 64-bit unsigned integer) by one.


\begin{figure}[t]
\centering
\adjustbox{varwidth=\linewidth,scale=0.8}{%
\lstset{language=C,
    basicstyle=\ttfamily,
    keywordstyle=\bfseries,
    showstringspaces=false,
    frame=single,
    morekeywords={include, printf}
}
\begin{lstlisting}
#include <stdint.h>
uint64_t incr(uint64_t i) { return i + 1; }
\end{lstlisting}
\lstset{
    language={[RISCV]Assembler},  
    basicstyle=\ttfamily,
    keywordstyle=\bfseries,
    showstringspaces=false,
    frame=single,
}
\vspace{-0.5em}
\begin{lstlisting}
incr:     file format elf64-littleriscv
Disassembly of section .text:
0000000000010488 <incr>:
   10488:	00150513          	addi	a0,a0,1
   1048c:	00008067          	ret
\end{lstlisting}
\lstset{
    language=SML,  
    basicstyle=\ttfamily,
    keywordstyle=\bfseries,
    showstringspaces=false,
    frame=single,
}
\vspace{-0.5em}
\begin{lstlisting}
<| bb_label := BL_Address_HC
    (Imm64 0x10488w) "00150513 (addi a0,a0,1)";
   bb_statements :=
    [BStmt_Assign (BVar "x10" (BType_Imm Bit64))
     (BExp_BinExp BIExp_Plus
      (BExp_Den (BVar "x10" (BType_Imm Bit64)))
      (BExp_Const (Imm64 1w)))];
   bb_last_statement :=
    BStmt_Jmp (BLE_Label (BL_Address (Imm64 0x1048Cw))) |>
\end{lstlisting}
}
\caption{Example program that increments a 64-bit unsigned integer represented in C (top), corresponding RISC-V disassembly (middle) and BIR abstract syntax (bottom).} \label{fig:ex-prog}
\end{figure}


\paragraph{Compilation.}
Our toolchain and workflow does not require program source code, only a RISC-V (RV64G) binary. However, we expect many users to perform binary analysis on programs for which they have source code, which enables control over compiler optimizations which can affect the size of the binary.

\paragraph{Disassembly.} 
We disassemble RV64G binaries using  \texttt{gnu-objdump} to enable further processing. The middle box in Figure~\ref{fig:ex-prog} shows our example's disassembly. 

\paragraph{Lifting.}
Our lifting tool parses the disassembly using specifications of program bounds (address of initial instruction and address after final instruction to lift) and translates the indicated input into a BIR program inside HOL4. Formally, a BIR program is a list of \emph{blocks} containing statements in abstract syntax, as shown at the bottom of Figure~\ref{fig:ex-prog}. Lifting also produces a HOL4 theorem expressing, roughly, that the BIR program simulates the corresponding RISC-V program---this theorem is what enables later backlifting of contracts from BIR to RISC-V.
\paragraph{RISC-V specification.}
We express RISC-V binary contracts as pairs of pre- and postconditions in HOL4 using the machine state defined by the L3 model, as illustrated at the top of Figure~\ref{fig:ex-rv-spec}, where \texttt{a0} is the standardized application binary interface (ABI) name of register \texttt{x10}.

\begin{figure}[t]
\centering
\adjustbox{varwidth=\linewidth,scale=0.8}{%
\lstset{
    language=SML,  
    basicstyle=\ttfamily,
    keywordstyle=\bfseries,
    showstringspaces=false,
    frame=single,
}
\begin{lstlisting}
Definition riscv_incr_pre_def: (* RISC-V precondition via L3 *)
 riscv_incr_pre (pre_x10:word64)(m:riscv_state) : bool =
 (m.c_gpr m.procID 10w = pre_x10)
End
Definition riscv_incr_post_def: (* RISC-V postcondition via L3 *)
 riscv_incr_post (pre_x10:word64) (m:riscv_state) : bool =
 (m.c_gpr m.procID 10w = pre_x10 + 1w)
End
\end{lstlisting}\vspace{-0.5em}
\begin{lstlisting}
Definition bir_incr_pre_def: (* BIR precondition *)
 bir_incr_pre (pre_x10:word64) : bir_exp_t = BExp_BinPred
  BIExp_Equal (BExp_Den (BVar "x10" (BType_Imm Bit64))) 
   (BExp_Const (Imm64 pre_x10))
End
Definition bir_incr_post_def: (* BIR postcondition *)
 bir_incr_post (pre_x10:word64) : bir_exp_t = BExp_BinPred
  BIExp_Equal (BExp_Den (BVar "x10" (BType_Imm Bit64)))
    (BExp_BinExp BIExp_Plus (BExp_Const (Imm64 pre_x10))
     (BExp_Const (Imm64 1w)))
End
\end{lstlisting}\vspace{-0.5em}
\begin{lstlisting}
Theorem bir_cont_incr: (* BIR contract via symbolic execution*)
 bir_cont bir_incr_prog bir_exp_true
  (BL_Address (Imm64 incr_init_addr))
  {BL_Address (Imm64 incr_end_addr)} {} (bir_incr_pre pre_x10)
  (\l. if l = BL_Address (Imm64 incr_end_addr)
     then bir_incr_post pre_x10 else bir_exp_false)
\end{lstlisting}\vspace{-0.5em}
\begin{lstlisting}
Theorem riscv_cont_incr: (* RISC-V contract via backlifting *)
 riscv_cont bir_incr_progbin incr_init_addr {incr_end_addr}
  (riscv_incr_pre pre_x10) (riscv_incr_post pre_x10)
\end{lstlisting}
}
\caption{Pre- and postconditions and theorems for program in Figure~\ref{fig:ex-prog}.} \label{fig:ex-rv-spec}
\vspace{-1em}
\end{figure}

\paragraph{BIR specification.}
To enable symbolic execution, we (manually) translate RISC-V pre- and postconditions to BIR, exemplified below the top box in Figure~\ref{fig:ex-rv-spec}. 

\paragraph{Symbolic execution, transfer, and backlifting.}
Once the BIR specification is set up, we can symbolically execute the BIR program. This process uses the BIR precondition as a path condition and produces a general symbolic execution theorem.
We then use this theorem to automatically prove a BIR contract and finally a RISC-V contract, as shown in the lowermost parts of Figure~\ref{fig:ex-rv-spec}. 

\section{Techniques and Implementation}
Our toolchain is implemented in the HolBA library for HOL4, i.e., it is implemented in Standard ML (SML). We summarize the implemented techniques.

\subsection{Architecture, lifting, and contracts}
\begin{figure}[t]
\centering
\includegraphics[width=0.8\textwidth]{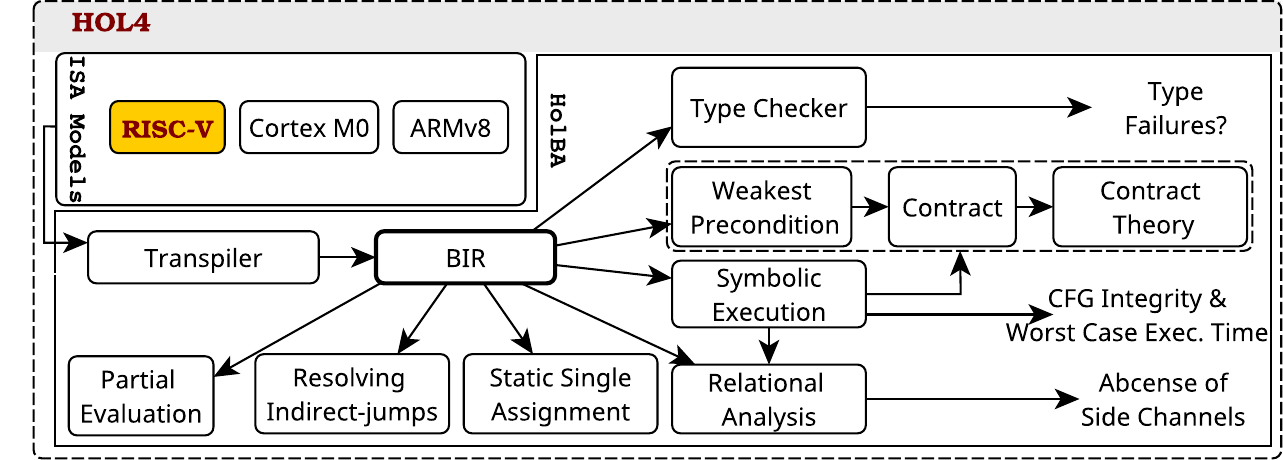}
\caption{Architecture of HolBA, a theorem-prover-based binary analysis platform. The figure illustrates HolBA's modular design, highlighting key components.} 
\label{fig:holba}

\end{figure}

HolBA (Figure~\ref{fig:holba}) is built based on the HOL4 theorem prover~\cite{hol4} and the L3 specification language~\cite{foxl3modelsweb} to facilitate analysis of binary code in an architecture-agnostic manner using BIR.
It ensures soundness of the analysis by formally certifying the transpilation of binary code into BIR, which involves translating machine code instructions into BIR snippets while  preserving their semantics. This is achieved using a proof-producing transpiler built on HOL4 ISA specifications that generates (bi)simulation theorems at the transition system level to ensure that the binary program's behavior is faithfully mirrored in BIR~\cite{dblp:conf/sbmf/meterelg17}.

HolBA provides both a HOL4 interface and an SML interface for most operations to facilitate its integration and usage. Its backend provides tools for symbolic execution, relational analysis, and contract verification, where contracts are \textit{Hoare triples}---tailored for binary programs with unstructured control flows, including those with multiple endpoints---to establish correctness of programs. HolBA's weakest precondition generation automates the verification of loop-free BIR code by deriving verification conditions that can be discharged using SMT solvers~\cite{Lundberg2020}. Symbolic execution complements this by tracking program states symbolically to explore execution paths of the given program and propagation of path conditions~\cite{proofproducingsymbexecution}. These techniques are tightly integrated with a Hoare-style logic to modularize rigorous correctness proofs. 

\subsection{Symbolic execution}
In order to automate verification, we rely on forward symbolic execution.
Compared to backward techniques, like weakest precondition generation, forward symbolic execution eases the propagation of constants and precondition constraints, enabling pruning of infeasible paths and optimization of context-sensitive operations.
This is especially useful in the context of binary code analysis, where
unstructured code can lead to many potential paths due to indirect jumps, and unstructured data may lead to memory aliasing.
For example, when returning from a callee function, the caller return address is known and propagated from the caller function; or when reading from the memory, the latest store operation can be ignored as the two accesses are disjunct under the path condition.

Following previous work~\cite{proofproducingsymbexecution},
a symbolic state $\bar s$ consists of a path condition and a mapping from variables to symbolic expressions, which in turn refer to symbols.
An interpretation $H$ maps symbols to values, and we say that a symbolic state $\bar s$ matches a concrete state $s$ when its evaluation under the interpretation $H$ renders the path condition true and the mapped variables as the values in $s$, written as $H(\bar s) = s$.
In order to deal with unstructured binary code, the symbolic executor relies on the concept of 
symbolic structures, $\bar s \rightarrow_L \{ \bar s_1 \dots \bar s_n\}$, which denotes that
executions encountering only the program labels $L$ and starting from a concrete state that is matched by the initial symbolic state $\bar s$ via an interpretation, eventually reach a concrete state that is matched by at least one of the symbolic states $\{ \bar s_1 \dots \bar s_n\}$ via the same interpretation.

The symbolic executor exploits HOL4 meta-programming and consists of two parts. The first component is a set of formally verified 
inference rules that allow to establish symbolic structures for individual instructions, compose them sequentially, do case analyses,
prune infeasible states, rename symbols, introduce abbreviations in the path conditions, simplify expressions in the states,
strengthen the path condition of the initial symbolic state, or weaken the path condition of a final state.
The second component is an ML procedure that uses the inference rules to analyze a complete program fragment.
This procedure uses several heuristics to choose which rules should be applied and how parameters should be instantiated, e.g. which expressions should be abbreviated,
which cases should be introduced.
Furthermore, this process is proof producing, meaning it uses the HOL4 system to justify the rule soundness theorems it chooses and composes them to produce the result as theorem.
To automatically prove a) infeasible path conditions for pruning symbolic states, b) expression simplification conditions, and c) contractual postcondition entailment for all final states in the symbolic structure,
we produce corresponding conditions using the SMTLIB standard and check them with an external SMT solver---currently Z3.
Finally, we use the BIR contract and lifter theorems to produce a RISC-V contract theorem.



\section{Binary Verification Case Studies}

To validate the usefulness of our toolchain and demonstrate that it can be used to verify real-world binaries, we performed case studies for two high-assurance system software domains: cryptography and operating systems.

\subsection{The ChaCha20 stream cipher}
ChaCha20 is a stream cipher that performs twenty rounds of the ChaCha algorithm as defined by Bernstein~\cite{Bernstein2008}. We obtained a ChaCha20 binary by compiling a reference implementation in C using gcc with optimization level -O1, yielding RISC-V disassembly with 847 instructions. Lifting of the dissassembly and performing symbolic execution on it took a few minutes on a modern machine.

We focused on specifying and verifying the byte encryption loop body of the ChaCha20 binary, taking inspiration from similar specifications in HacSpec~\cite{bhargavan2018hacspec}. More specifically, we define a ChaCha round as a transformation of 16 variables holding unsigned 32-bit words (array in the source code), which is then used as input to the next round. 
Figure~\ref{fig:chacha-spec} shows our definition of a ChaCha \emph{quarter round} transforming the state, which we then used as a building block to specify the byte encryption loop body as a so-called \emph{column round} followed by a \emph{diagonal round}~\cite{rfc7539}. In the figure, \texttt{??} is the HOL4 word XOR operation, \lstinline[basicstyle=\ttfamily]{<<~} is word left shift, \lstinline[basicstyle=\ttfamily]{>>>~} is word right shift, and \lstinline[basicstyle=\ttfamily]{#<<~} is word rotation to the left.

\begin{figure}[t]
\centering
\adjustbox{varwidth=\linewidth,scale=0.8}{%
\lstset{
    language=SML,  
    basicstyle=\ttfamily,
    keywordstyle=\bfseries,
    showstringspaces=false,
    frame=single,
}
\begin{lstlisting}
Definition chacha_line_fst_def:
 chacha_line_fst (pre_a:word32) (pre_b:word32) = pre_a + pre_b
End
Definition chacha_line_snd_def:
 chacha_line_snd pre_a pre_d (s:word32) : word32 =
 ((pre_a ?? pre_d) <<~ s) || ((pre_a ?? pre_d) >>>~ (32w - s))
End (* equivalent to (pre_a ?? pre_d) #<<~ s when s <=+ 31w *)
Definition chacha_quarter_round_def:
 chacha_quarter_round pre_a pre_b pre_c pre_d =
  let a = pre_a; b = pre_b; c = pre_c; d = pre_d;
  a = chacha_line_fst a b; d = chacha_line_snd a d 16w;
  c = chacha_line_fst c d; b = chacha_line_snd c b 12w;
  a = chacha_line_fst a b; d = chacha_line_snd a d 8w;
  c = chacha_line_fst c d; b = chacha_line_snd c b 7w 
  in (a,b,c,d)
End
\end{lstlisting}
}
\caption{High-level ChaCha20 quarter round specification in HOL4.} \label{fig:chacha-spec}

\end{figure}

Note that our verification of ChaCha byte encryption only establishes the connection between a high-level functional specification in HOL4 and how the main loop body of the RISC-V binary behaves according to L3. That is, we did not prove any non-functional (security) properties of the ChaCha20 binary.

\subsection{RISC-V separation kernel}
S3K is an open source capability-based separation kernel targeting embedded RISC-V systems~\cite{s3k}. Practically, S3K manages a collection of 
\emph{user processes} and ensures that they adhere to given restrictions on execution time (slices), memory access, and interprocess communication. 
%
%
We specified and verified two key handwritten RISC-V assembly routines from S3K:
\begin{description}
\item[\texttt{trap\_entry}:] Entered when the user process is interrupted or an exeception is encountered. Loads a pointer to the process's process control block (PCB) from the special \texttt{mscratch} RISC-V register, then stores the user process's context (general purpose registers) to the PCB.

\item[\texttt{trap\_exit}:] Entered when the kernel resumes a user process. Loads the user process's context from the PCB (pointed at by register \texttt{x10}), replacing register values with the values from the PCB.
\end{description}
We obtained disassembly for the routines from the S3K author, consisting of 81 instructions that takes less than a minute to lift and symbolically execute. After adjusting lifting for special RISC-V instructions such as \texttt{csrrw}, we expressed routine contracts in terms of assignment of registers to memory locations (\texttt{trap\_entry}), and assignments of memory locations to registers (\texttt{trap\_exit}); Figure~\ref{fig:kernel} shows a fragment of the (RISC-V) postcondition of \texttt{trap\_entry}. The S3K author manually validated our specifications, and we then translated them to BIR manually and performed the verification separately for each routine.

\begin{figure}[t]
\centering
\adjustbox{varwidth=\linewidth,scale=0.85}{%
\lstset{
    language=SML,  
    basicstyle=\ttfamily,
    keywordstyle=\bfseries,
    showstringspaces=false,
    frame=single,
}
\begin{lstlisting}[language=SML]
Definition riscv_kernel_trap_entry_post_def:
riscv_kernel_trap_entry_post (pre_x1:word64) (pre_mhartid:word64)
(pre_mscratch:word64) (pre_mepc:word64) (*..*) (m:riscv_state):=
m.c_gpr m.procID 2w = (* adjusted stack pointer in x2 register *)
 (0xFFFFFFFFFFFFFFFFw * (pre_mhartid <<~ 10w)) + 0x80006400w /\
riscv_mem_load_dword m.MEM8 (pre_mscratch+8w) = pre_mepc /\
riscv_mem_load_dword m.MEM8 (pre_mscratch+16w) = pre_x1 (*..*)
End
\end{lstlisting}}
\caption{Fragment of postcondition for \texttt{trap\_entry} in S3K disassembly.} 
\label{fig:kernel}
\end{figure}


\section{Symbolic Execution Evaluation}
\label{sec:eval}
To investigate the scalability of our proof-producing symbolic execution, we ran it on a collection of dissassembled RISC-V binaries and measured the running time; the results are shown in Table~\ref{tbl:symbexec}. To give an indication of the binary size, the table lists the number of RISC-V instructions; this is close to but not the same as the number instructions symbolically executed, e.g., we often skip \texttt{ret}.

\begin{table}[t]
\centering
\adjustbox{varwidth=\linewidth,scale=0.9}{%
\begin{tabular}{lcll}
\textbf{Program} & \textbf{\#Instr.} & \textbf{Description} & \textbf{Time}\\
\hline
aes-unopt   & 637 & AES cipher, optimization -O0                 & 67.65s\\
aes         & 264 & AES cipher, optimization -O1                 & 5.52s\\
motor       & 120 & motor driver, nested functions and branching & 3.71s\\
swap        & 8   & swap memory location content                 & 0.19s\\
isqrt       & 15  & integer square root                          & 0.11s\\
mod2        & 4   & modulo two                                   & 0.03s\\
incr        & 4   & increment variable                           & 0.03s
\end{tabular}
}
\vspace{0.3cm}
\caption{Evaluated programs and their symbolic execution running times.} \label{tbl:symbexec}
\vspace{-0.5cm}
\end{table}

Per the table, symbolic execution of BIR programs corresponding to disassembly with hundreds of instructions generally takes at most a few minutes. The running time increases with both the number of BIR statements (corresponding to ISA instructions) and the ``\textit{store complexity}'' of the program. The latter refers to how much information needs to be stored in the symbolic state during execution. In many cryptographic routines, all the output bits depend on all the input bits, yielding a large symbolic state as in AES.
In the unoptimized binary, these expressions are replicated several times in the symbolic memory.
Note that AES is only symbolically executed from initialization until the end of the first round.

\section{Conclusions and Future Work}
\label{sec:conclusion}
We presented a workflow for verifying RISC-V binaries inside the HOL4 theorem prover using the HolBA library for binary analysis~\cite{holbasymbexec}. The workflow is based on orchestrating and augmenting previously disparate parts of HolBA, including lifting of RISC-V machine code to the BIR intermediate language, Hoare-style binary contracts, and forward symbolic execution of BIR. In particular, we automated generation of BIR contract theorems from symbolic execution theorems.

We demonstrated the toolchain on case studies from high-assurance system software domains, highlighting that binary verification is mainly useful for ensuring that binaries follow high-level behavioral specifications. 
Future work includes automated translation of (restricted) RISC-V contracts to BIR contracts, which may also enable an interface to the toolchain outside of HOL4. 

\bibliographystyle{splncs04}
\bibliography{bib}
\end{document}